# Spatio-temporal propagation of traffic jams

# in urban traffic networks

Date: May 19, 2017

Jiang Yinan[1,2], Kang Rui[1,2], Li Daqing*[1,2], Guo Shengmin[3,4], Shlomo Havlin[5]

1. School of Reliability and Systems Engineering, Beihang University, Beijing 100191, China
2. Science and Technology on Reliability and Environmental Engineering Laboratory, Beijing 100191, China
3. State Key Laboratory of Software Development Environment, Beihang University, Beijing 100191, China
4. Beijing PalmGo Infotech Co., Ltd, Beijing 100191, China
5. Department of Physics, Bar Ilan University, Ramat Gan 5290002, Israel

* Correspondence and requests for materials should be addressed to Li Daqing (daqingl@buaa.edu.cn)

## Abstract

Since the first reported traffic jam about a century ago, traffic congestion has been intensively studied with various methods ranging from macroscopic to microscopic viewpoint. However, due to the population growth and fast civilization, traffic congestion has become significantly worse not only leading to economic losses, but also causes environment damages. Without understanding of jams spatio-temporal propagation behavior in a city, it is impossible to develop efficient mitigation strategies to control and improve city traffic. Although some progress has been made in recent studies based on available traffic data regarding general features of traffic, the understanding of the spatio-temporal propagation of traffic jams in urban traffic is still unclear. Here we study the spatio-temporal propagation behavior of traffic jams based on collected empirical traffic data in big cities. We developed a method to identify influential jam centers and find that jams spread radially from multiple jam centers with a range of velocities. Our findings may help to predict and even control the traffic jam propagation, which could be helpful for the development of future autonomous driving technology and intelligent transportation system.

# Introduction

The frequency and intensity of traffic jam have been increasing in recent years worldwide concern due to high impact on the economy and pollution[1-4]. Although the earliest research about traffic management started in the 1930s[5], modern cities from all over the world suffer from growing traffic congestion. Records show that in 471 U.S. urban areas, traffic jams caused 6.9 billion vehicle-hours of delay in 2014, which is almost three times from 1982. Moreover, traffic congestion results in 3.1 billion gallons of wasted fuel and for a loss of $160 billion[6]. Besides, environmental pollution caused by traffic congestion becomes an increasingly serious health problem. Approximately one-third of fine particulate matter (PM2.5) in urban areas in the U.S. results from vehicles[7, 8]. The frequent acceleration and braking during traffic jams exacerbate the air pollution. A recent study by Harvard School of Public Health suggests that the air pollution from traffic congestion in the 83 largest urban areas of the U.S. contributes to more than 2,200 premature deaths annually[9].

Studies of traffic jams and their propagation have been performed since 1950s. A bottleneck in a traffic network is considered as a source of a traffic jam[10]. Initially congestion occurred in a bottleneck, and from there, the traffic jam may propagate to its surroundings. Therefore, identification of traffic jam centers that propagate is essential prerequisite for understanding the propagation behavior. In simulations and empirical studies, the bottleneck is usually identified according to the macroscopic physical quantities such as traffic flow[11], velocity[12] and occupancy rate[13], combining with models considering the microcosmic characteristic of traffic flow and the topology of traffic network[14]. Recently, evolving critical bottlenecks in real-time data of city road traffic have been identified in real traffic by characterizing the organization process of traffic as "traffic percolation"[15].

To better understand the traffic congestion mechanism, we analyze here the spatio-temporal propagation pattern of traffic jams in a large city. Local congestion may spread in a city traffic network, as a result of multiple factors such as routing selection[16], driving behavior[17], road capacities[18, 19] and even weather effects[20, 21].

In such a complex-open system, the propagation behavior of traffic jams is hard to predict and control. But in the future Internet of Vehicles, one of the critical techniques required is to calculate precisely in real time, predict effectively and avoid spatio-temporal evolution of traffic jams. This could be based on the huge amounts of traffic data, which include the location, the velocity, the routing, the status information of each vehicle and the interactive information between them. For reaching this target we wish here to understand the spatio-temporal propagation behavior of traffic jams not only in single lanes of highways but also from the perspective of a city traffic network.

Different types of traffic flow models and simulations have been proposed in an attempt to understand the propagation of traffic jams[22-26]. These include microscopic car-following and cellular automata models[27-29], mesoscopic gas-kinetic-based models and macroscopic fluid mechanics models[30-32]. In parallel, some empirical studies provided realistic support for the above-mentioned models with respect to the phase transition mechanism and the phase transition points during formation and dissipation of traffic jams[33-37]. By analyzing the spatio-temporal statistics of density or velocity on highways, some empirical studies find that the jams move along a lane with an almost stationary structure and the estimated velocity of the moving jam's downstream front is about 15 km/h[38-41]. This has been observed in cities in USA, UK and Germany. Recently, the properties of jam propagation have been studied by analyzing the spatial pattern of traffic jams and long-range correlation features of traffic jams has been found in daily city traffic with correlations decaying slowly with distance[42]. The dynamical behavior of propagation has been studied in a model based on cascading overload failures in spatially embedded traffic networks and an approximately constant propagation velocity has been found[43].

Currently, most research on the origin and propagation of traffic jams use traffic flow models under assumptions about traffic network topology and traffic assignment, which may deviate from the real situation. Moreover, most studies mainly focus on single lanes or roads, but not characterizing the jam propagation behavior in a scale of a city traffic network. Furthermore, while the existing studies focus on the static statistical properties of traffic jams, the question of spatio-temporal propagation behavior is still open.

In this paper, we collect and analyze real-time road velocity records in two major cities in China, Beijing and Shenzhen (see SI). Influential jam centers that cause jam propagation have been identified and a decay of jam strength with the distance from centers has been found to follow an approximate linear relation. We also find indications of a wave phenomenon in the propagation of traffic jams and estimated velocities of jam waves are obtained.

## Results

The collected real-time traffic data span the urban area of Beijing (within 5$^{th}$ Ring Highway), which include the whole road network with 52,968 roads. Our data of traffic with velocities at each road covers 14 working days of 24 hours at 1-min resolution, in October 2015. The velocity of each road $v_i(t)$ varies during a day according to real-time traffic. Considering that the traffic capacity varies over different roads, we define traffic jams through a relative velocity threshold $q$. For each road at time $t$, if the ratio between its current velocity $v_i(t)$ and its maximal velocity measured during that day falls below the threshold $q$, it is considered as a failure (jam). Otherwise, it is regarded as normal state. Considering practical and general criterion of traffic jams, we assume the value of 0.25 for $q$. In fact, similar features and findings on the propagation of traffic jams are found for other values of $q$. The results of two other values of $q$ (0.20 and 0.30) are demonstrated and compared in SI (Fig. 1 and Fig. 6).

A regional traffic congestion usually roots in congestion occurred on a segment of a road. Varying with the road conditions, the "origin points" or jam centers of traffic jams are flickering on urban traffic networks during a day. In order to study the statistical characteristics, possible regularities and the spatio-temporal propagation of these jam centers, we analyze the congestion degree of each road segment and their distributions.

The statistics of flow in each road segment could help to find the hidden heterogeneity of traffic congestion and identify the critical segments during jam propagation. Here, we propose two measures to quantify (i) the congestion degree of the road via congestion frequency and (ii) congestion duration. The first is the occurrence probability of congestion (*occp*), which measures the fraction of road congestion during a given period of time,

$$occp = \frac{t_F}{T}. \qquad (1)$$

Here $T$ is the total window time and $t_F$ is the number of congested instants (of 1 min resolution) in this window. In Fig. 1(a), we demonstrate and compare the changes in state in five typical roads with different values of *occp* (1 represent congestion). For the roads with high values of occp, the congested states tend to be gathered together in time series rather than scatter randomly in the form of short segments. The probability density distributions of *occp* for different 2-hour time periods during a day for all roads are shown in Fig. 1(b) (see also Fig. 2 in SI for the city of Shenzhen). The occurrence distributions of traffic jams seem to decay approximately as a power-law in particular during rush hours. The slope varies slightly with time and reaches its minimum at evening rush hours 17:00-19:00 of about 1.2, which suggests scaling laws and long-term decay of the congestion probability of road segments.

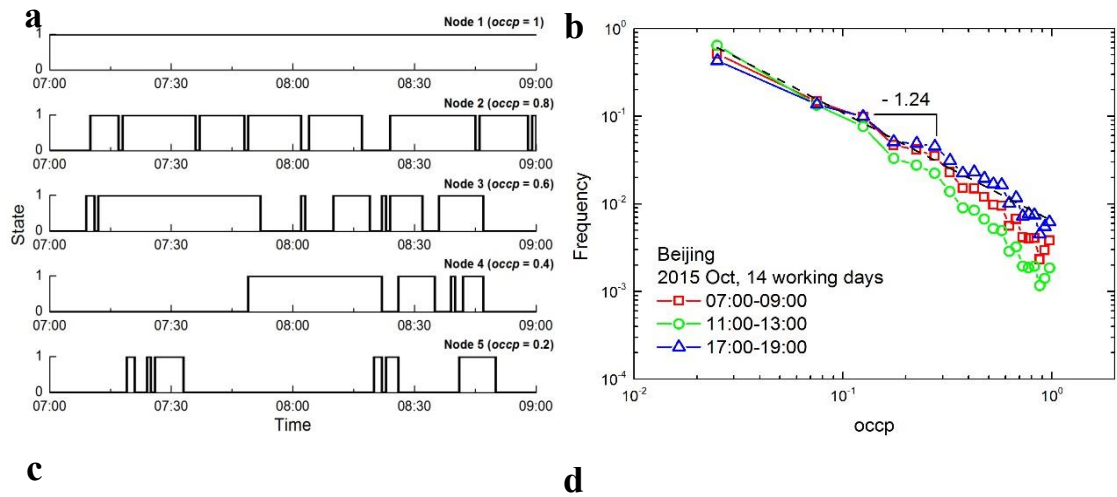

a  b

c  d

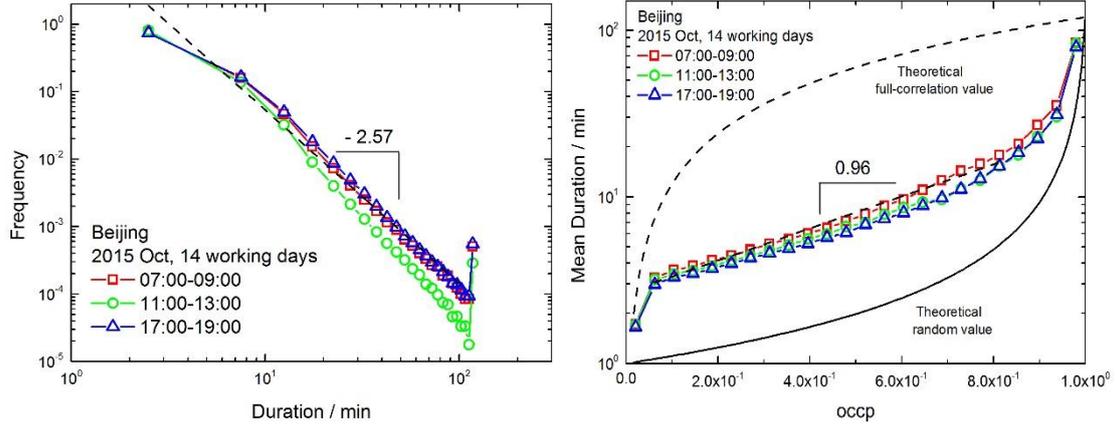

**Figure 1 Identification of jam centers in the traffic network of Beijing.** (a) Demonstration of the occupation state (1 represent jam) of 5 typical roads during morning rush hours 07:00-09:00. The roads with highest value of *occp* and mean duration are identified as jam centers (here is Node 1). (b) Distribution of occurrence probability of congestion (*occp*) for three typical 2-hour time periods during a day. (c) Distribution of congestion duration for the three typical 2-hour time periods. (d) Mean duration as a function of *occp*. The squares, circles and triangles are results of real traffic data. The solid curve is the theoretical value of random control and the dashed curve is the theoretical value of full-correlation cases (i.e., all jams are in a single cluster in the time series).

In order to characterize the temporal properties of traffic congestion of each road segment, we calculate duration of congestion which measures the persistent congested state. That is how long a road is trapped in the congested state after smooth traffic. Fig. 1(c) shows the distribution of congestion duration of all roads in urban traffic network for three typical 2-hour time periods during a day (results for the city of Shenzhen can be seen in SI). The results suggest a power-law distribution with the exponent close to 2.6.

Slow decay and spatial heterogeneity features of both distributions of congestion probability and persistent time suggest strong temporal correlation between traffic jam instances and possibly indicate the existence of jam centers in an urban traffic network. In Fig. 1(d) we analyze the relation between mean duration and *occp*. Mean duration of a road is the averaged values of all congested durations in a period. The mean duration value reflects the expected dissipation time of a traffic jam on the road. Indeed, we can see in Fig. 1(d) that the mean duration of real traffic data falls between the control random case and the full-correlation situation, which indicates the strong temporal correlations between jam instances to clamp together (3-4 times more than random). More results are presented in SI.

Combining this autocorrelation feature with the findings about slow decay distributions of occurrence probability and congestion duration, the roads with both highest value of *occp* and longest congested duration are chosen to be potential jam centers that propagate in space. We assume that, for a given time window (e.g. 7:00-9:00 a.m.) the *occp* of the roads to be considered as jam centers is 1 or close to 1. It means that jam centers are the roads which are congested all the time during the period and we test how much they continuously influence the traffic conditions of surrounding roads. Remarkably, there are often several such possible roads. It confirms to the concurrency of traffic jams in megacities, because big cities usually have several traffic regions, commercial areas of high population density that are susceptible to traffic congestion.

Jam centers do not only hinder the traffic flow through them, but in many cases also influence surrounding area. They are usually determined jointly by several factors such as commuter flow, traffic capacity and the local topology structure around them. Once being jammed, the traffic flow through them would slow down and vehicles may choose to bypass them. In this way, jam may spread from centers. Fig. 2(a) illustrates multiple jam centers and the congestion status around them (Fig. 2(b)) in urban traffic network of Beijing. As mentioned above, potential jam centers are identified based on their high rank of *occp* and marked as red solid dots. The congestion status of roads around them are demonstrated in Fig. 2(b) as circles with different shades. The lighter the shade, the faster the traffic flow through it. Here we use failure rate, *fr*, to quantify the evolution of congestion strength at a given time. The failure rate $fr$ is defined as:

$$fr(r_c, t) = \frac{F(r_c, t)}{M(r_c)}. \tag{2}$$

Here, $F(r_c, t)$ is the number of roads that become congested at time $t$ at location $r_c$ (distance from the jam center), and $M(r_c)$ is the total number of roads at $r_c$. The failure rate, $fr$, measures the fraction of new congested road at current instant which is fluctuating due to the variation of traffic speed. The shades of circles in Fig. 2(b) become lighter through the decrease of $fr$ with $r_c$. The values of $fr$ might be influenced by jam centers and reflect the spatio-temporal evolution of jam strength from the centers.

Thus, quantifying the effect of jam centers on their surrounding neighborhood is critical for deepening our understanding of the propagation behavior of traffic jams. Here, for urban traffic network we study the relationship between traffic centers and the spreading based on the dependence of $fr$ on distance from the center.

By calculating $fr$ variation at different spatial locations around centers and different times, we find an approximate linear decay of $fr$ with the increase of $r_c$ (see in Fig. 2(c)):

$$fr(r_c, t) = k_t r_c + c_t \ . \tag{3}$$

Here $k_t$ is the slope of $fr(r_c, t)$ function and $c_t$ is the corresponding intercept. This linear decay is relatively stable within certain time period during a day. The results shown in Fig. 2(c) are for several 20-min segments of morning peak-hour. During morning rush hours, the slope $k$ varies approximately in the range from -0.0005 to -0.0015. Similar results for other two time periods (off-peak and evening peak hours) can be seen in SI. For non-centers, the linear decay feature does not exist. The results for randomly selected jam centers can be seen in SI.

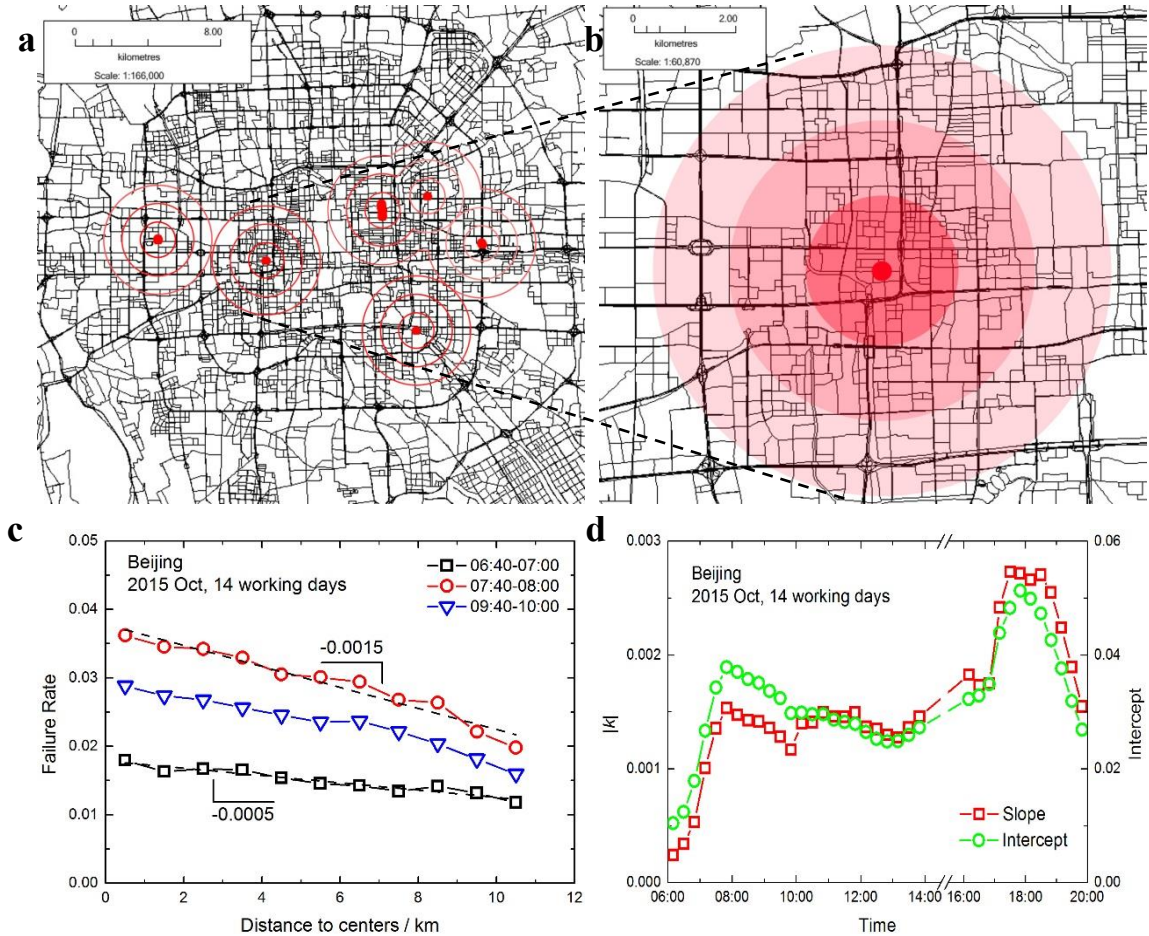

**Figure 2 Decay of jam strength from the center.** (a) Illustration of multiple jam centers in traffic network of Beijing. Traffic data is from 6:00 a.m. to 10:00 a.m. on Monday October 26[th], 2015. (b) Illustration of a typical congestion status around one typical jam center. It is observed on 7:30-7:40 a.m., which is a critical moment toward a widespread congestion, on Monday October 26[th], 2015. (c) Failure rate $fr$ as a function of distance from jam centers in three different 20-min periods of morning rush hours. Each curve is an average of 14 working days. (d) Fitting slopes $k$ of failure rate curves in (c) and the intercept $c$ at different times of a day.

Based on the findings of linear decay during the spatial propagation of traffic jams, we notice that the linear slope $k$ varies with time $t$. The larger the slope $k$, the faster the decay of traffic jams. So, we attempt to find out the characteristics of the linear slopes $k$ and the corresponding intercepts $c$ with varying of time $t$ throughout the day. The results of $k$ and $c$ averaged over 14 working days are shown in Fig. 2(d). Note that, an almost simultaneous change trend is found for $k$ and $c$. The behavior in Fig.

2(d) could be understood since $c$ measures the congestion level near jam centers and $k$ measures the rate of decay. Note that the failure rate reduces to a background level far away from jam centers (here the influential distance is about 10 km) and the background failure rate changes less with time. It means that the more serious of the traffic jams in centers, the faster they decay with distance. As we can see, the slope $k$ is extremely small in the early morning, which can be explained that the overall situation of traffic network is in a good condition. For the rest of the day, the variation of $k$ can be divided into two stages. One corresponds to evening peak period. The slope $k$ increases as time evolves towards evening rush hours and reaches its peak value at the most congested time, which corresponds to about 17:50 in the evening. Our suggestion is that the overall traffic condition is the joint outcome of the severity of congestion in jam centers and the linear decay slope $k$. Therefore, the observation of maximum $k$ indicates that at evening peak-hour the worst traffic condition occurred in critical jam centers represented by both, high values of intercepts and the fastest decay effect. The other stage includes morning and noon, the slope $k$ fluctuates slightly in an intermediate range, indicating an almost regular traffic condition.

Considering the spatial heterogeneity of traffic jams, we further investigate the propagation of jams strength from *single* centers in different locations. For this purpose, we regard jam centers located closely in space (within 2 km) as a single jam cluster. As seen in Fig. 3(a) and (b), there are both 7 jam clusters during 06:00-10:00 on Oct 26[th], 2015, and 16:00-20:00 on Oct 28[th], 2015 in the traffic network of Beijing, respectively. Each jam cluster is shown by a different color. Significant differences have been found between these jam clusters when measuring the spatial evolution of the jam strength. For each period of time, two extreme examples of jam propagation are selected and compared with random control, see Fig. 3(c) and (d). As we can see, some jam centers (such as Cluster 1 in Fig. 3(a) and Fig. 3(b)) have high impact on jam propagation in their surrounding area which lead to a much higher failure rate compared to the random background, see Fig. 3(c) and (d). Similar to that found in the global traffic network, the influence strength of single clusters decay approximately linear with the increase of distance from the center. However, we can see that the influence on jam propagation of some jam centers (such as Cluster 7 in Fig. 3(a)) is much weaker or even absent. The failure rate around these centers changes little with distance and the values are close to those of random control.

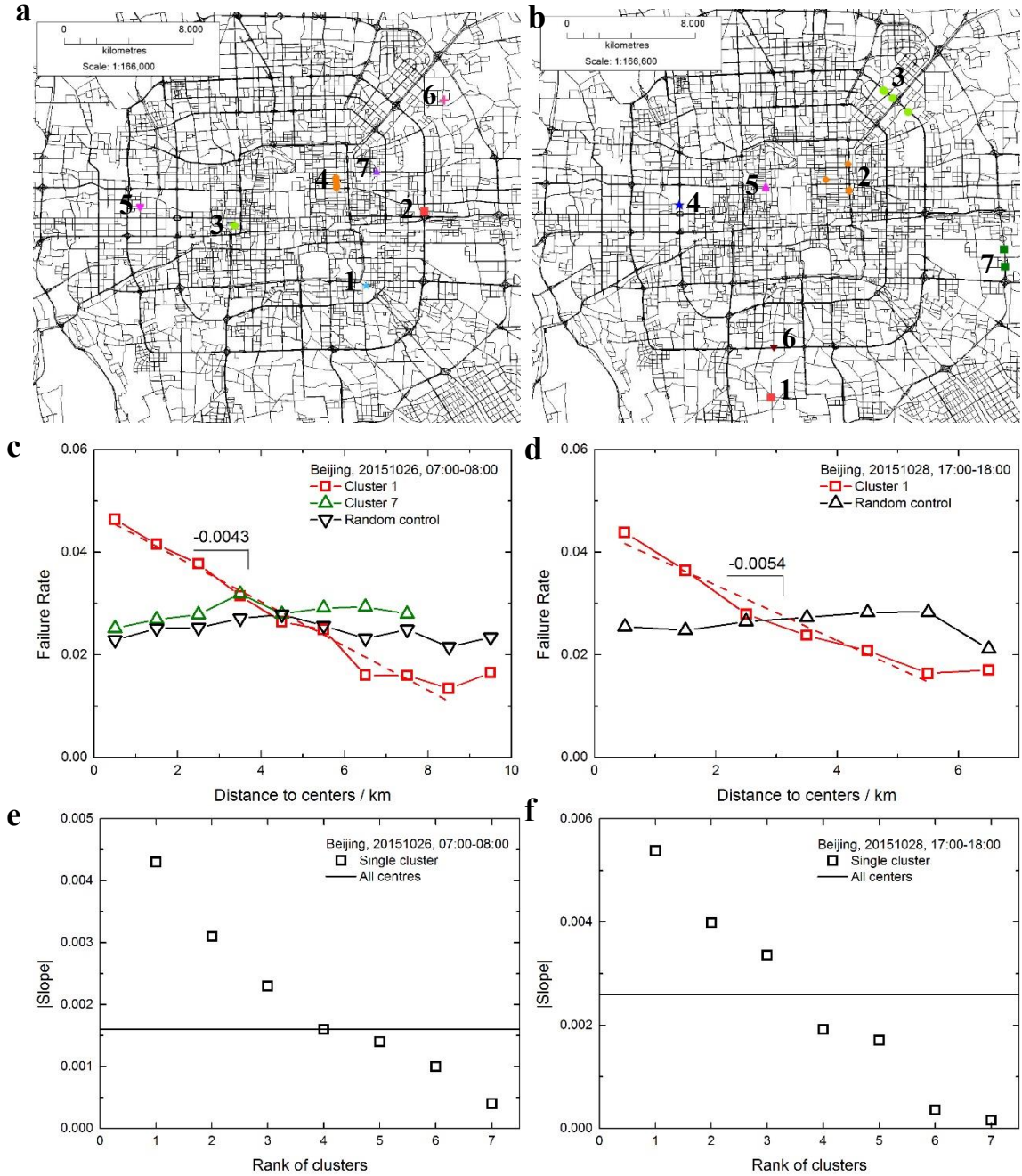

**Figure 3 The influence on jam propagation of different clusters of jam centers.** (a) and (b) Illustration of clusters of jam centers in the traffic network of Beijing during morning 06:00-10:00 on October 26th, 2015, and evening 16:00-20:00 on October 28th, 2015, respectively. (c) and (d) Failure rate $fr$ as a function of distance from centers for different clusters of jam centers. (e) and (f) Comparison of slopes $k$ of failure rates vs. distance for different clusters and that of the averaged global slope.

Therefore, when considering single different jam centers, there are some real centers that collectively dominate the spatial propagation of traffic jams. To find these critical jam centers, we compare the slope of the $fr$ function of each jam cluster with the average value of the global situation. As seen in Fig. 3(e) and (f), the clusters of these two periods of time have been ranked according to their slopes. It is seen that only about half of the jam clusters seem to influence and propagate to their surrounding area. While for the other half of clusters, sustained congestions are confined to themselves without impacting their neighborhood by spatial spreading of jams. In order to visually display the location and the spatial distribution of these influential jam clusters, their rankings are labeled in Fig. 3(a) and (b). Thus, our results suggest that counter intuitively, whether a jam cluster is a real spreading center or not is a dynamic feature and might have no special relation to its location in the traffic network. We also find that the locations of jam centers vary from day to day (see Fig. 10 in SI).

Based on the overall general feature of static spatial evolution of jam propagation from the center, we further investigate the dynamic characteristics during the propagation process. The temporal fluctuations of $fr$ at different locations are shown in Fig. 4(a). Each curve corresponds to a given distance from jam centers which is marked in the right. From bottom to top, the distance $r_c$ is increasing. As we can see from Fig. 4(a), the broad trends of $fr$ indicate that the jam strength at a given place has a periodic variation while $fr$ fluctuates up and down almost every minute. This occurs for every spatial distance range cell. Besides the temporal dynamic, we study the spatial dynamical propagation of jams by calculating $fr$ as a function of distance from centers $r_c$ at a given time $t$ (see SI). Similar fluctuations of $fr$ appear in space where above the overall background of decay, $fr$ goes up and down with increasing $r_c$ at any time, see Fig. 4(c).

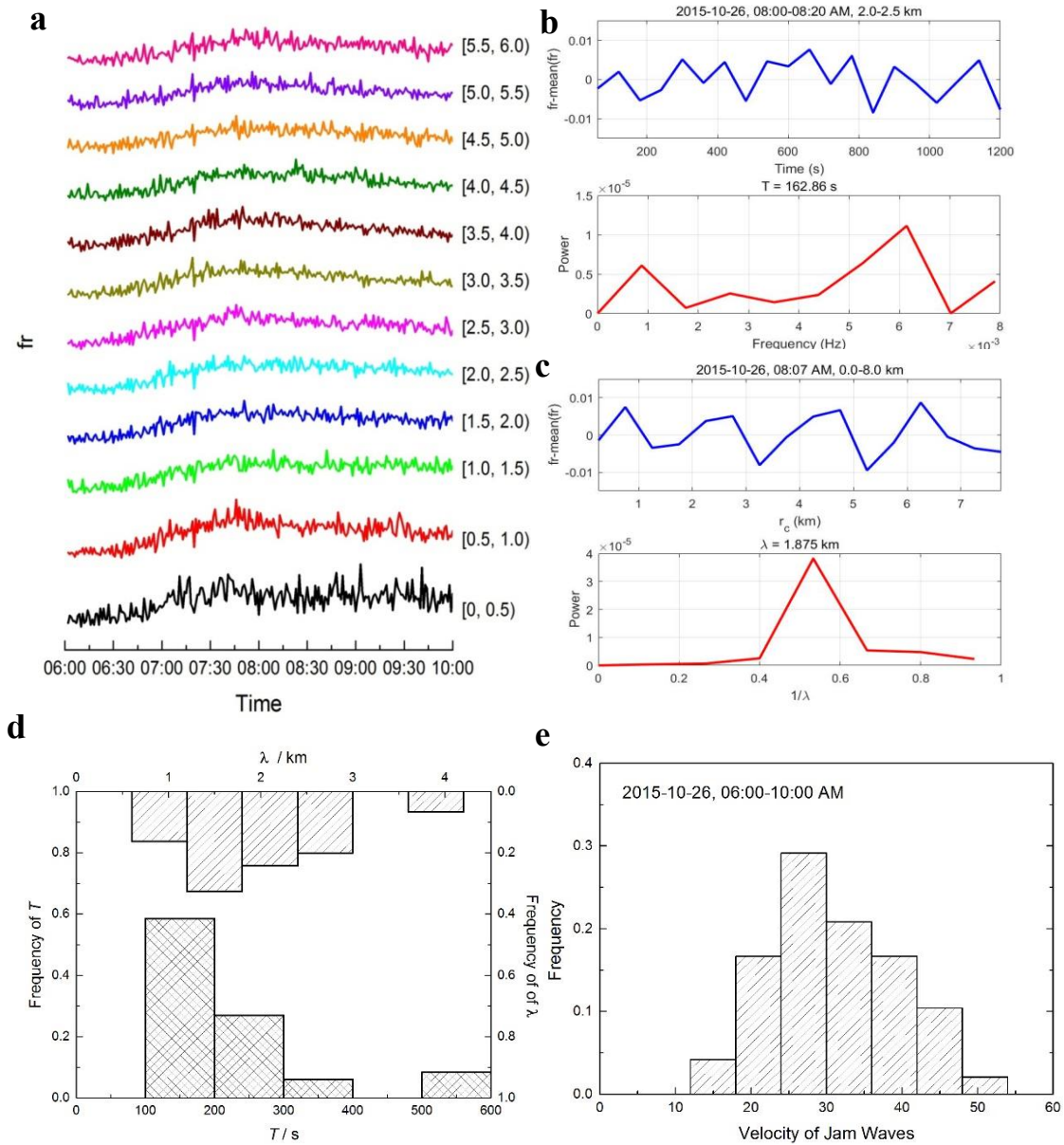

**Figure 4 Velocities of jam waves.** (a) $fr$ as a function of time at different $r_c$. The resolution is 1 minute. (b) $(fr - \overline{fr})$ as a function of time and the power spectrum for the range of $r_c$ in 2.0-2.5km. The period calculated based on the power spectrum is 163s. (c) $(fr - \overline{fr})$ as a function of $r_c$ and the power spectrum for the time at 08:07 AM. The wavelength calculated based on the power spectrum is 1.9km. (d) and (e) Distributions of periods and wavelengths of jam waves. (f) Distribution of velocities of jam waves in Beijing.

The variations presented in both time dimension and space dimension suggest a wave phenomenon in the propagation of traffic jams. This means that the severe and sustained local traffic congestion in center may radiate out in the form of jam waves.

The attenuation characteristics of jam waves in traffic network can explain the linear decay of $fr$ with spatial distance from centers, where the amplitude of the wave is decreasing with distance. For a specific spatial location, $fr$ can demonstrate periodic peaks corresponding to the period of jam waves. To study the wave phenomenon of $fr$ in time dimension, we analyze the power spectrum of $fr$ vs. $t$ curves in Fig. 4(a). First, to filter the component of overall trend, the mean value of $fr$ of each 30-min time window is subtracted from $fr$ within the same time window. As an example, the temporal fluctuations of $(fr - \overline{fr})$ for $r_c$ between 2 km and 2.5 km are shown in the upper part of Fig. 4(b). Then the estimated period can be calculated by spectrum analysis shown in the lower part of Fig. 4(b). Analogously, the spatial fluctuations at a given instant and the corresponding wavelength are demonstrated in Fig. 4(c).

The distributions of periods and wavelengths of jam waves that spread in the traffic network are shown in Fig. 4(d). As can be seen, over 30% of jam waves have a wavelength of 1.5km (most probable) and over 70% of wavelengths are in the range of 0.6-2.4km. And the most frequent period of jam waves is in the range between 100 seconds and 200 seconds. In order to investigate the impact of traffic lights on the propagation of traffic jams, we measure the distributions of periods and distances of traffic lights in the main intersections of two areas in Beijing. As shown in SI Fig. 15, over 80% of periods of traffic lights are in the range of 100-200 seconds, no matter in rush hours (7:00-8:00 a.m. and 17:00-18:00 p.m.) or non-rush hours. And this applies to different areas of the Beijing traffic network. Since the most probable range of periods of traffic lights is almost same as that found in jam waves, it seems that there is certain relation between these two quantities. In contrast, for distances between main traffic lights, the distributions show shorter ranges than those found in our jam waves (see SI Fig. 16). This may be caused by several continuous green lights that drivers can cross without stopping.

The indications of a wave phenomenon in the propagation of traffic jams suggest the possibility to represent the evolution of congestion strength at any given location and time by a function $fr(r_c, t)$ based on the measured quantities. In analogy with general wave equations, we assume that the function of jam waves can be approximated as,

$$fr(r_c, t) = A(r_c, t) \left|\cos(\frac{t}{T} - \frac{r_c}{\lambda} + \varphi_0)\right|. \tag{4}$$

Here $T$ is jam wave period and $\lambda$ is jam wavelength. $A(r_c, t)$ is the amplitude of the jam wave located at $r_c$ at time $t$. Combined with the linear decay suggested in Eq. (3), the function of jam waves can be written as

$$fr(r_c, t) = (k_t r_c + c_t) \left|\cos\left(\frac{t}{T} - \frac{r_c}{\lambda} + \varphi_0\right)\right|. \tag{5}$$

The propagation velocity of a jam wave is an important quantity because it could help in predicting when and where the congestion effect would arrive from jam centers. Here, it could be calculated based on the wavelength and wave period information obtained above. Then from the spatio-temporal variation of $fr$, we can calculate the current velocity of the jam wave anywhere and anytime through dividing current wavelength by wave period. The distributions of velocities of jam waves spreading in traffic network of Beijing are shown in Fig. 4(e). It can be seen that the velocity of jam waves in the morning hours 06:00-10:00 a.m. in Beijing is about 30±10 km/h. It is nearly twice as that found in one-dimensional lanes on a highway [38, 40]. Distributions for other time periods in Beijing and results for Shenzhen can be seen in SI. This quantity is of significance for mitigation and prediction of traffic jams because it can tell us where and when would the spreading jams be.

## Conclusion

The spatio-temporal propagation behavior of traffic jam originated from jam centers in urban traffic networks has been studied. The linear decay of jam strength from identified jam centers has been found as a function of distance from a center. Furthermore, we reveal real jam centers that jointly dominate the propagation of traffic jams by zooming in to study the spreading of single centers. By analyzing the spatio-temporal dynamic fluctuations of traffic jams, the form of jam waves has been approximated during propagation process and estimated spreading velocities of jam waves have been obtained.

Based on these essential features of jam propagation, we can predict and might control jam conditions in the city traffic networks. These findings could be useful for optimizing transportation efficiency and improving traffic control strategies. Furthermore, it might be useful for the future Intelligent Transportation System(ITS).

## Methods

### Data description

The dataset analyzed in the manuscript is the sampling velocity data collected by floating cars in two major cities in China, Beijing and Shenzhen. The real-time traffic data of Beijing and Shenzhen covers 14 working days in October 2015, with resolution of 1-minute segments for 24 hours during a day. Our data spans the urban area of the cities which includes the typical and general traffic congestion situations. It covers the road network with 52,968 roads in Beijing and 22,248 roads in Shenzhen.

### Data preprocessing

Due to the inevitable defect during the sampling of floating cars, there are no real-time velocity data for some roads at a specific moment. The data missing rate is about 25% overall. Therefore, we need to fill up the missing velocity data at each time in order to complete the traffic condition in the whole road network. Here, we apply a space compensation method, which regards the averaged velocity of all inflow and outflow neighboring roads of a missing data of a road as its current velocity. By an iterative process, the compensation of missing velocity data can be completed for each instant.

Another important issue is the establishment of failure/jam criterion. As we know, the road condition and capacity are diverse for different road levels. So, the failure/jam criterion should vary according to specific situations of different roads. Relative velocity ratio is taken as a measurement related to the criterion. It is the ratio between the current velocity $v_i(t)$ of road $i$ at time $t$ and its maximal velocity measured for the whole day. This quantify makes comparable velocities for different roads and a relative velocity threshold q then proposed to judge the state of each road.

**Identification of jam centers in traffic network**

As we have shown in Fig. 1 in the manuscript, the probability density distributions of *occp* (occurrence probability of congestion) and duration of congestion seem to decay approximately as a power law during rush hours. The spatial heterogeneity in congested probabilities of roads and expected congestion time of roads suggest the existence of jam centers in an urban traffic network. Moreover, the mean duration of a jam is found to increase approximately (for low *occp*) in a form of power law function of *occp* with an exponent close to 1 (almost liner), significantly higher compared to random. These observations indicate the strong temporal correlations between jam instances to clamp together.

Therefore, we choose the roads with both highest value of *occp* and longest congested duration as potential jam centers. We assume that, for a given time window (e.g. 7:00-9:00 a.m.) the *occp* of the roads to be considered as jam centers is 1 or close to 1. In fact, for 2-hour time periods the number of roads that selected as jam centers is 15 and 10 in Beijing and Shenzhen, respectively.

Further, we identify the critical jam centers as those that dominate the spatial propagation of traffic jams. By investigating the dependence of failure rate *fr* with distance from each single center clusters, we find significant difference of the spatial evolution of the jam strength between single clusters. Then the center clusters that have high impact on jam propagation are identified as critical jam centers.

# References

1. Helbing, D., et al., *Self-Organized Pedestrian Crowd Dynamics: Experiments, Simulations, and Design Solutions.* Transportation Science, 2005. **39**(1): p. 1-24.
2. Petri, G., et al., *Entangled communities and spatial synchronization lead to criticality in urban traffic.* Scientific Reports, 2013. **3**(5): p. 1224-1230.
3. Orosz, G., R.E. Wilson, and G. Stépán, *Traffic jams: dynamics and control.* Philosophical Transactions of the Royal Society A Mathematical Physical & Engineering Sciences, 2010. **368**(1928): p. 4455-79.
4. Nagatani, T., *Traffic jam at adjustable tollgates controlled by line length.* Physica A Statistical Mechanics & Its Applications, 2015. **442**: p. 131-136.
5. Greenshields, B.D., *A Study in Highway Capacity, Highway Research Board.* 1935.
6. Schrank, D.L., et al., *2015 Urban Mobility Scorecard.* 2015.
7. Rizzo, M.J. and P.A. Scheff, *Fine particulate source apportionment using data from the USEPA speciation trends network in Chicago, Illinois: Comparison of two source apportionment models.* Atmospheric Environment, 2007. **41**(29): p. 6276-6288.
8. Hammond, D.M., et al., *Sources of ambient fine particulate matter at two community sites in Detroit, Michigan.* Atmospheric Environment, 2008. **42**(4): p. 720-732.
9. Levy, J.I., J.J. Buonocore, and K.V. Stackelberg, *Evaluation of the public health impacts of traffic congestion: a health risk assessment.* Environmental Health, 2010. **9**(1): p. 1-12.
10. Wright, C. and P. Roberg, *The conceptual structure of traffic jams.* Transport Policy, 1998. **5**(1): p. 23-35.
11. Kerner, B.S., *Empirical macroscopic features of spatial-temporal traffic patterns at highway bottlenecks.* Physical Review E Statistical Nonlinear & Soft Matter Physics, 2002. **65**(4 Pt 2A): p. 046138.
12. Long, J.C., et al., *Urban traffic congestion propagation and bottleneck identification.* Science China Information Sciences, 2008. **51**(7): p. 948.
13. Menendez, M. and C.F. Daganzo, *Effects of HOV lanes on freeway bottlenecks.* Transportation Research Part B Methodological, 2007. **41**(8): p. 809-822.
14. Sun, H., et al., *Spatial distribution complexities of traffic congestion and bottlenecks in different network topologies.* Applied Mathematical Modelling, 2014. **38**(2): p. 496-505.
15. Li, D., et al., *Percolation transition in dynamical traffic network with evolving critical bottlenecks.* Proceedings of the National Academy of Sciences of the United States of America, 2015. **112**(3): p. 669-72.
16. S, Ç., A. Lima, and M.C. González, *Understanding congested travel in urban areas.* Nature Communications, 2016. **7**: p. 10793.
17. Chiabaut, N., L. Leclercq, and C. Buisson, *From heterogeneous drivers to macroscopic patterns in congestion.* Transportation Research Part B Methodological, 2010. **44**(2): p. 299-308.
18. Rajagopalan, S. and H.L. Yu, *Capacity planning with congestion effects.* European Journal of Operational Research, 2001. **134**(2): p. 365-377.
19. Osorio, C., *An analytic finite capacity queueing network model capturing the propagation of congestion and blocking.* European Journal of Operational Research, 2009. **196**(3): p. 996-1007.
20. Kwon, J., M. Mauch, and P. Varaiya, *The Components of Congestion: Delay from Incidents,*

*Special Events, Lane Closures, Potential Ramp Metering Gain, and Excess Demand, presented at 85 th Annual Meeting Transportation Research Board.* 2006.
21. Tanaka, K., T. Nagatani, and H. Hanaura, *Traffic congestion and dispersion in Hurricane evacuation.* Physica A Statistical Mechanics & Its Applications, 2007. **376**(1): p. 617-627.
22. Helbing, D., et al., *Micro- and macro-simulation of freeway traffic.* Mathematical & Computer Modelling, 2002. **35**(5–6): p. 517-547.
23. Nagatani, T., *The physics of traffic jams.* Reports on Progress in Physics, 2002. **65**(9): p. 1331-1386(56).
24. Kai, N., P. Wagner, and R. Woesler, *STILL FLOWING: APPROACHES TO TRAFFIC FLOW AND TRAFFIC JAM MODELING.* Operations Research, 2003. **51**(5): p. 681-710.
25. Wilson, R.E., *Mechanisms for spatio-temporal pattern formation in highway traffic models.* Philosophical Transactions of the Royal Society A Mathematical Physical & Engineering Sciences, 2008. **366**(1872): p. 2017-32.
26. Orosz, G., et al., *Exciting traffic jams: nonlinear phenomena behind traffic jam formation on highways.* Physical Review E, 2009. **80**(4 Pt 2): p. 046205.
27. Lighthill, M.J. and J.B. Whitham, *On Kinematic Waves: I. Flow movement in Long Rivers. II. A theory of Traffic Flow on Long Crowded Roads.* Pharmacology & Therapeutics, 1955. **53**(3): p. 275-354.
28. Nagatani, T., *Density waves in traffic flow.* Physical Review E Statistical Physics Plasmas Fluids & Related Interdisciplinary Topics, 2000. **61**(4 Pt A): p. 3564-70.
29. Gupta, A.K. and V.K. Katiyar, *Analyses of shock waves and jams in traffic flow.* Journal of Physics A General Physics, 2005. **38**(38): p. 4069-4083(15).
30. Sipahi, R. and S.I. Niculescu, *Stability of car following with human memory effects and automatic headway compensation.* Philosophical Transactions of the Royal Society A Mathematical Physical & Engineering Sciences, 2010. **368**(1928): p. 4563-83.
31. Wagner, P., *Fluid-dynamical and microscopic description of traffic flow: a data-driven comparison.* Philosophical Transactions of the Royal Society A Mathematical Physical & Engineering Sciences, 2010. **368**(1928): p. 4481-95.
32. Chandler, R.E. and E.W. Montroll, *Traffic Dynamics: Studies in Car Following.* Operations Research, 2012. **6**(2): p. 165-184.
33. Kerner, B.S. and H. Rehborn, *Experimental Properties of Phase Transitions in Traffic Flow.* Physical Review Letters, 1997. **79**(20): p. 4030-4033.
34. Treiber, M., A. Hennecke, and D. Helbing, *Congested traffic states in empirical observations and microscopic simulations.* Physical Review E Statistical Physics Plasmas Fluids & Related Interdisciplinary Topics, 2000. **62**(2 Pt A): p. 1805-24.
35. Popkov, V., et al., *Empirical evidence for a boundary-induced nonequilibrium phase transition.* Journal of Physics A General Physics, 2001. **volume 34**(34): p. L45-L52(8).
36. Treiber, M., A. Kesting, and D. Helbing, *Three-phase traffic theory and two-phase models with a fundamental diagram in the light of empirical stylized facts.* Transportation Research Part B Methodological, 2010. **44**(8–9): p. 983–1000.
37. Helbing, D., et al., *Theoretical vs. empirical classification and prediction of congested traffic states.* European Physical Journal B, 2009. **69**(4): p. 583-598.
38. Kerner, B.S. and H. Rehborn, *Experimental features and characteristics of traffic jams.* Physical Review E Statistical Physics Plasmas Fluids & Related Interdisciplinary Topics, 1996.

**53**(2): p. R1297-R1300.

39. Lieu, H., *The Physics of Traffic: Empirical Freeway Pattern Features, Engineering Applications, and Theory.* Physics Today, 2004. **58**(11): p. 54-56.

40. Rehborn, H., S.L. Klenov, and J. Palmer, *An empirical study of common traffic congestion features based on traffic data measured in the USA, the UK, and Germany.* Physica A Statistical Mechanics & Its Applications, 2011. **390**(23-24): p. 4466-4485.

41. Schönhof, M. and D. Helbing, *Empirical Features of Congested Traffic States and Their Implications for Traffic Modeling.* Transportation Science, 2007. **41**(2): p. 135-166.

42. Li, D., et al., *Spatial correlation analysis of cascading failures: congestions and blackouts.* Scientific Reports, 2014. **4**(4): p. 5381-5381.

43. Zhao, J., et al., *Spatio-temporal propagation of cascading overload failures in spatially embedded networks.* Nature Communications, 2016. **7**.

# Supplementary Figures

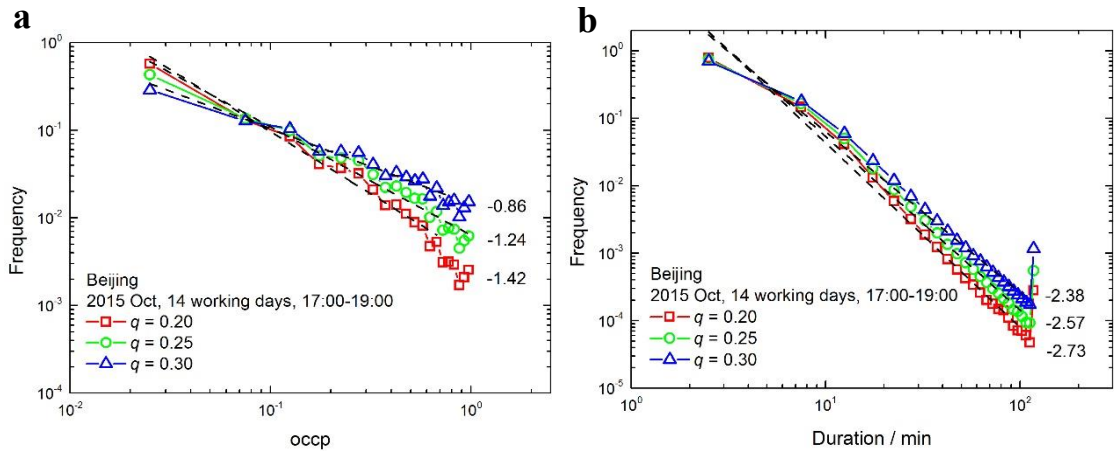

**Supplementary Figure 1: Effects of relative velocity threshold *q* on statistical characteristics of congestion.** (a) Distribution of occurrence probability of failure (*occp*) during 17:00-19:00, which corresponds to the most severe congestion of a day. (b) Distribution of congestion duration for the same 2-hour period. In these figures, three values of relative velocity threshold *q* are compared, including 0.20 (red squares), 0.25 (green circles) and 0.30 (blue up triangles).

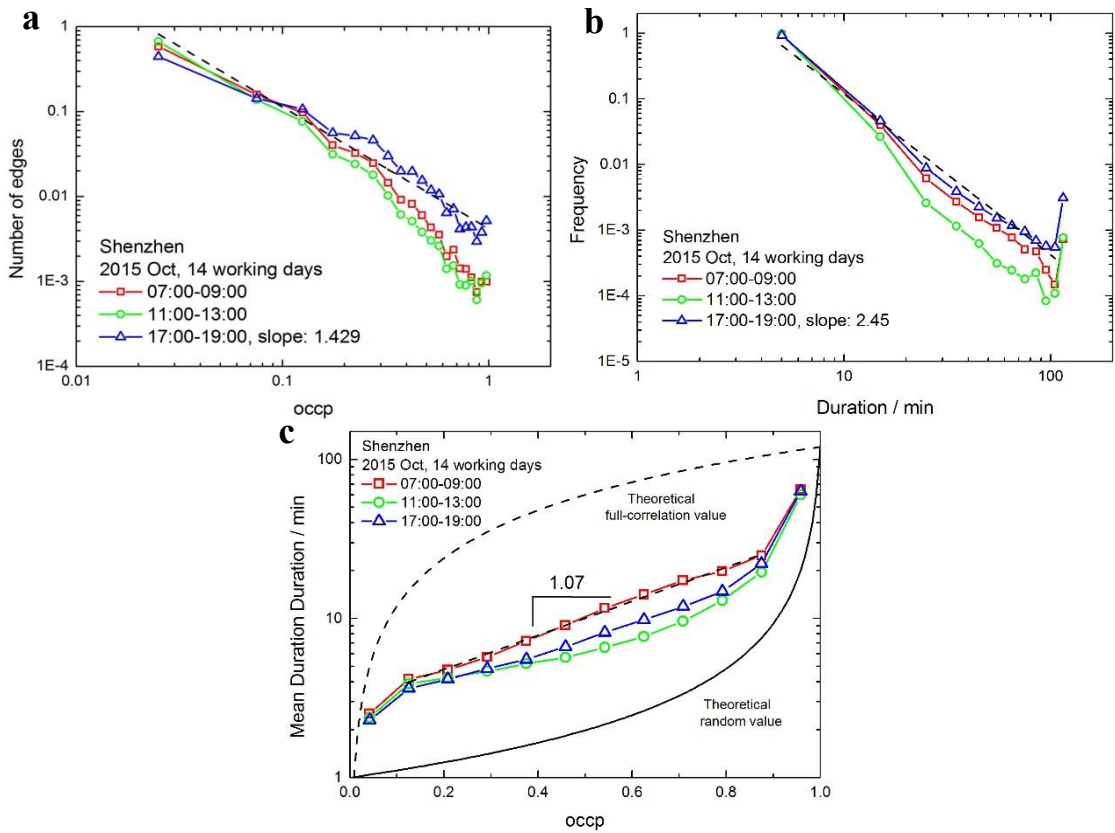

**Supplementary Figure 2: Identification of jam centers in traffic network of Shenzhen.** (a)

Distribution of occurrence probability of failure (*occp*) for different 2-hour time periods in Shenzhen, including 07:00-09:00 (red squares), 11:00-13:00 (green circles) and 17:00-19:00 (blue up triangles). (b) Distribution of congestion duration for the three typical 2-hour time periods. Results have been averaged over 14 working days. (c) Mean duration as a function of *occp*. The squares, circles and up triangles are results of real traffic data. The solid curve is the theoretical value of random cases and the dashed curve is theoretical value of full-correlation cases.

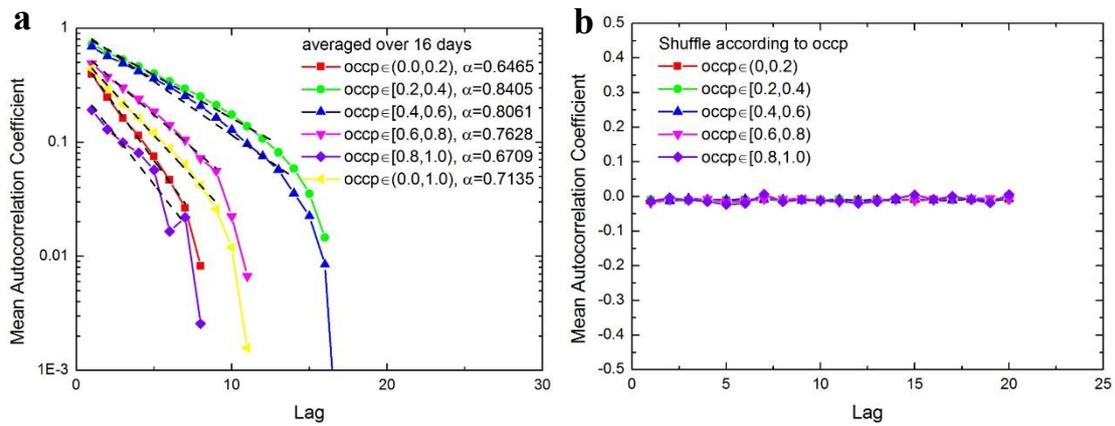

**Supplementary Figure 3: Temporal autocorrelation of status of roads.** (a) statistics for different intervals of *occp*. (b) Autocorrelation in random shuffling cases.

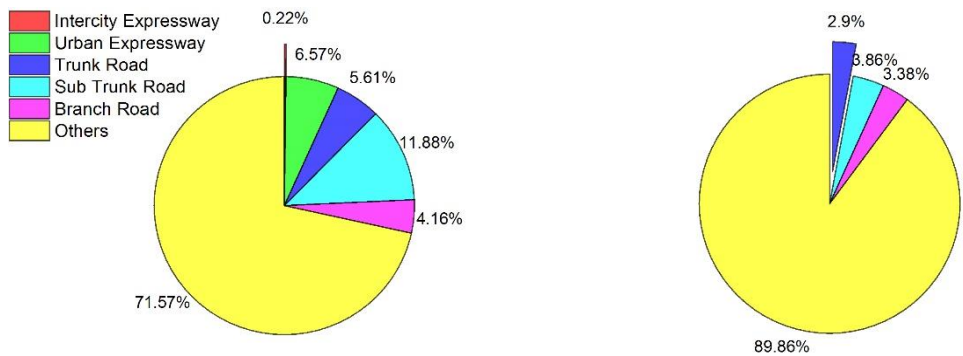

(a) All roads (within 5th Ring of Beijing)    (b) Jam centers (14 working days)

**Supplementary Figure 4: The static attributes of jam centers in traffic network of Beijing.**

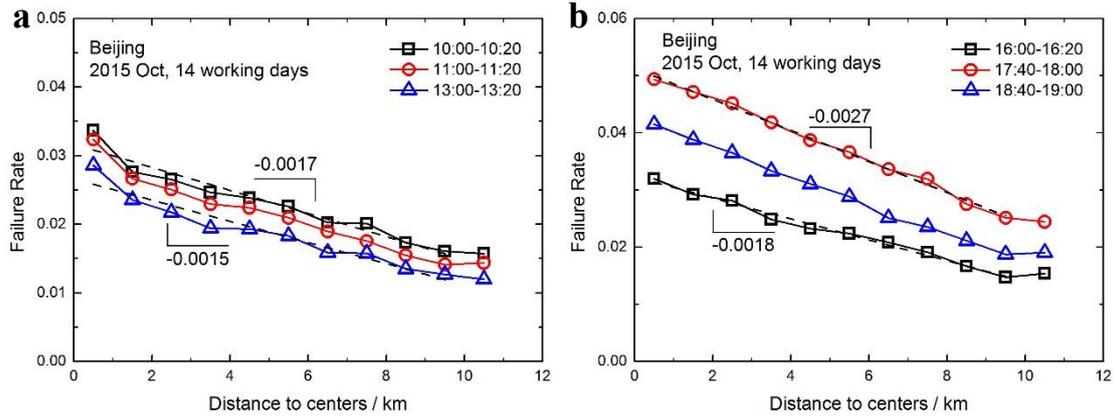

**Supplementary Figure 5: Decay of jam strength from the center in Beijing.** Failure rate $fr$ as a function of distance from jam centers in three different 20-min periods in (a) off-peak hours and (b) evening peak hours. Each curve is an average of 14 working days.

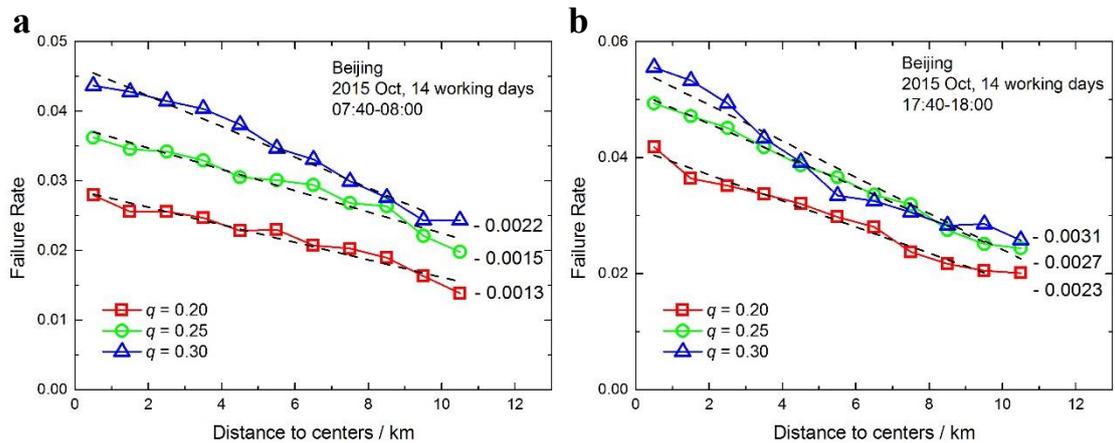

**Supplementary Figure 6: Effects of relative velocity threshold $q$ on decay of jam strength from the center.** Failure rate $fr$ as a function of distance from jam centers in two serious congestion 20-min periods in (a) morning peak hours and (b) evening peak hours. Three values of relative velocity threshold $q$ are compared, including 0.20 (red squares), 0.25 (green circles) and 0.30 (blue up triangles). Each curve is an average of 14 working days.

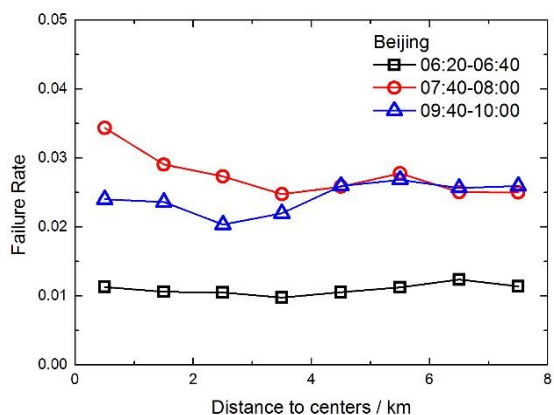

**Supplementary Figure 7: Jam strength around random jam centers of Beijing.** Failure rate $fr$ as a function of distance from randomly selected jam centers in three different 20-min periods of morning rush hours.

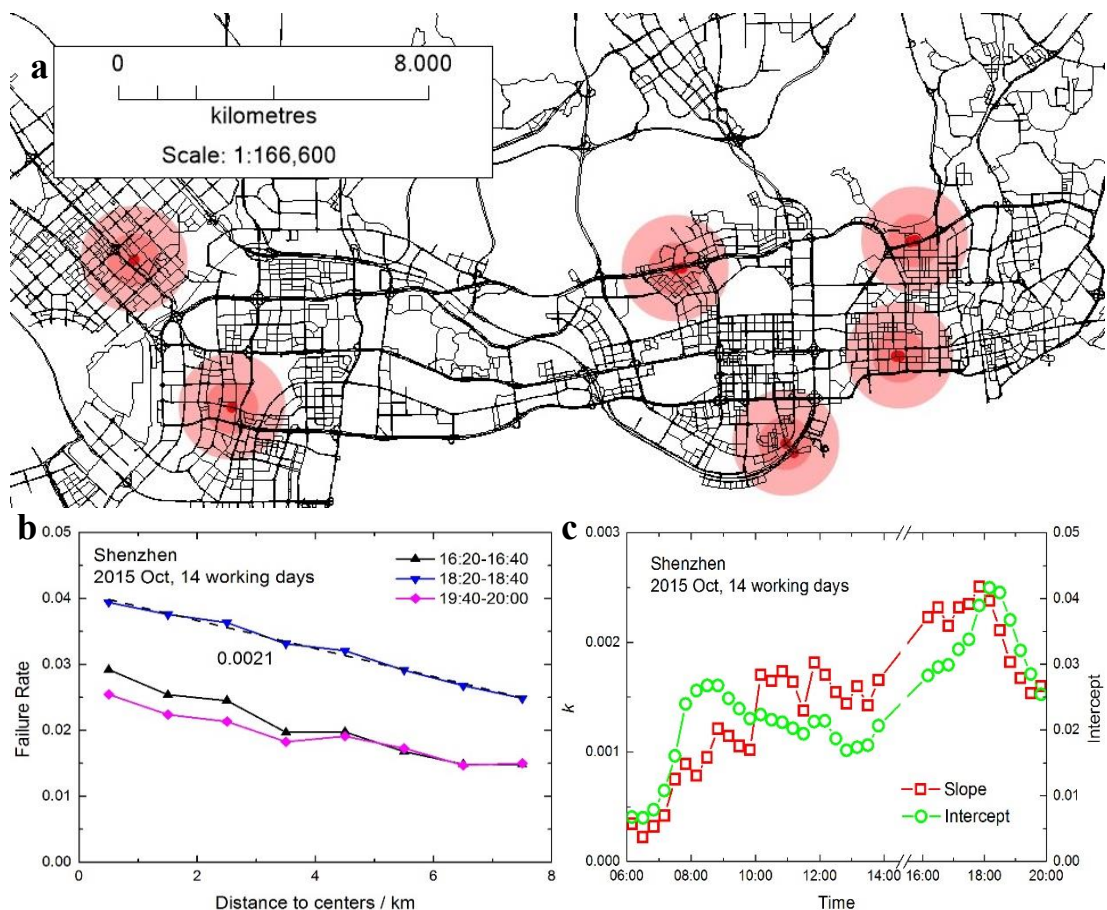

**Supplementary Figure 8: Decay of jam strength from the center in Shenzhen.** (a) Illustration of multiple jam centers in traffic network of Shenzhen. Traffic data is from 6:00 a.m. to 10:00 a.m. on Monday October 26th, 2015. (b) Failure rate $fr$ as a function of distance from jam centers in three different 20-min periods of evening rush hours. Each curve is an average of 14 working days.

(c) Fitting slopes $k$ of failure rate curves in (b) and the intercept $c$ at different times of a day.

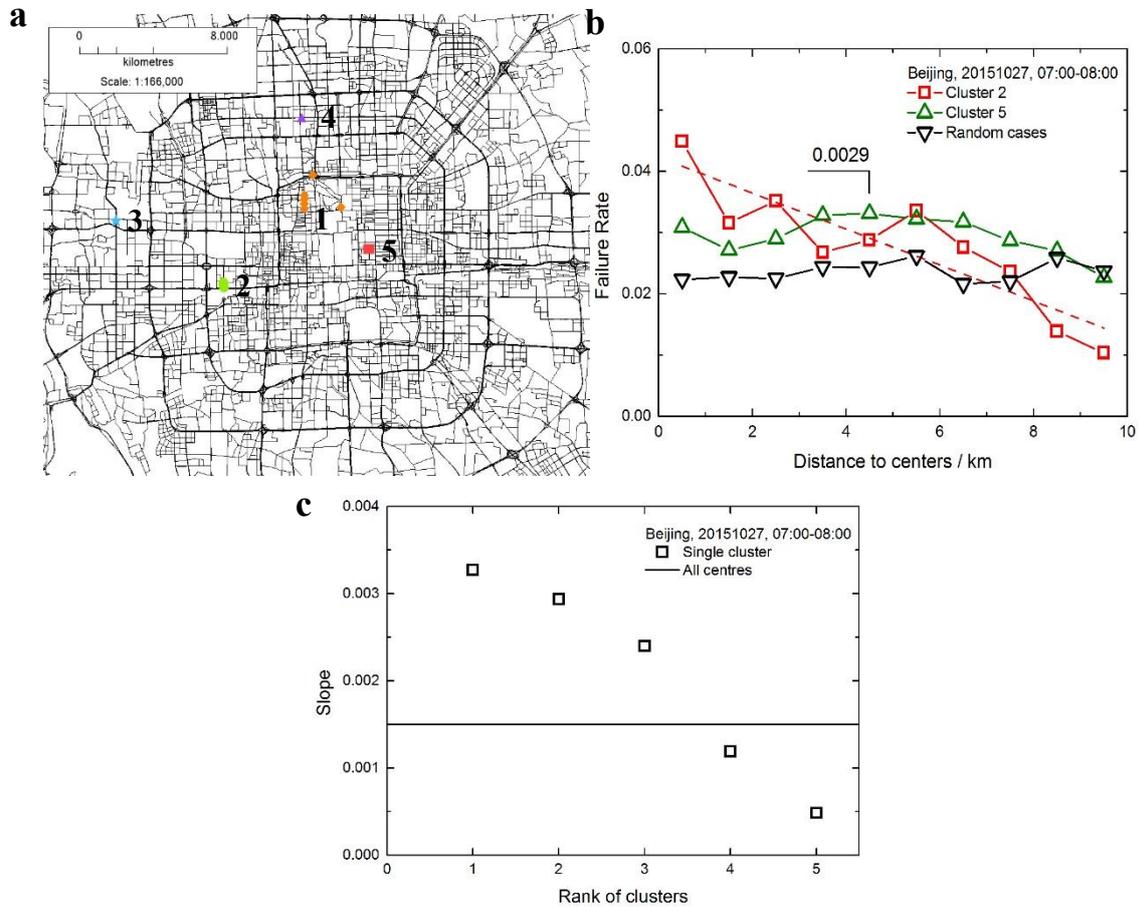

**Supplementary Figure 9: The influence on jam propagation of different clusters of jam centers.** (a) Illustration of clusters of jam centers in traffic network of Beijing during morning 06:00-10:00 on October 27[th], 2015. (b) Failure rate $fr$ as a function of distance from centers for different clusters of jam centers. (c) Comparison of slopes $k$ of failure rates for different clusters and that of the averaged global slope.

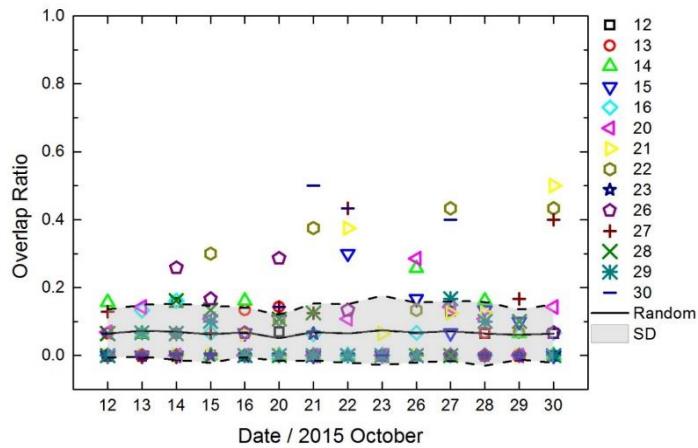

**Supplementary Figure 10: Overlap ratio of the locations of jam centers between different day pairs.** Overlap ratio is the ratio of *NOL* to *NC*. Here, *NOL* is the number of overlap centers of two days, which are defined as the centers whose minimal distance to the centers of another day is less than or equal to overlap radius (here is 1km) and vice versa. *NC* is the number of centers of two days (here is 30). The *x* axis is the date of October 2015, the black curve is the averaged result of random control. The random result is averaged over 1000 realizations and the light grey area corresponds to standard deviation.

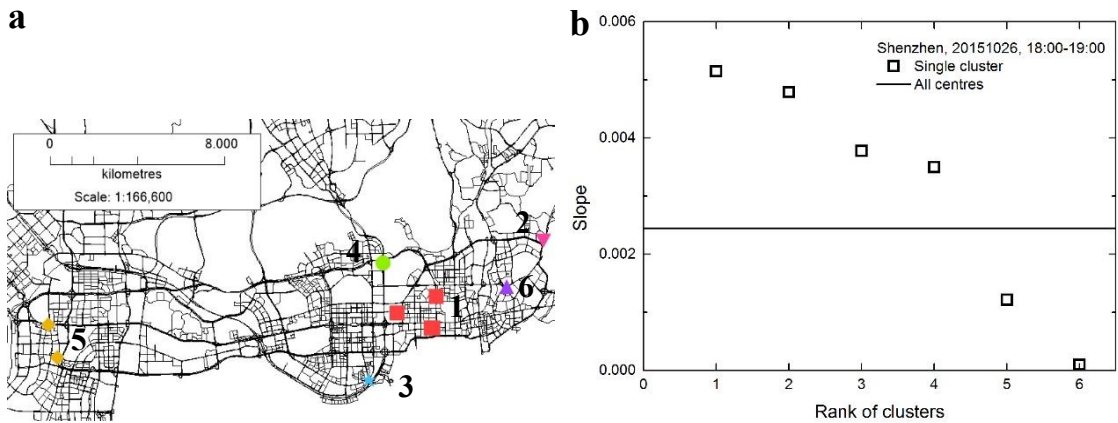

**Supplementary Figure 11: The influence on jam propagation of different clusters of jam centers in Shenzhen.** (a) Illustration of clusters of jam centers in traffic network of Shenzhen during evening 16:00-20:00 on October 26[th], 2015. (b) Comparison of slopes *k* of failure rates for different clusters and that of the averaged global slope.

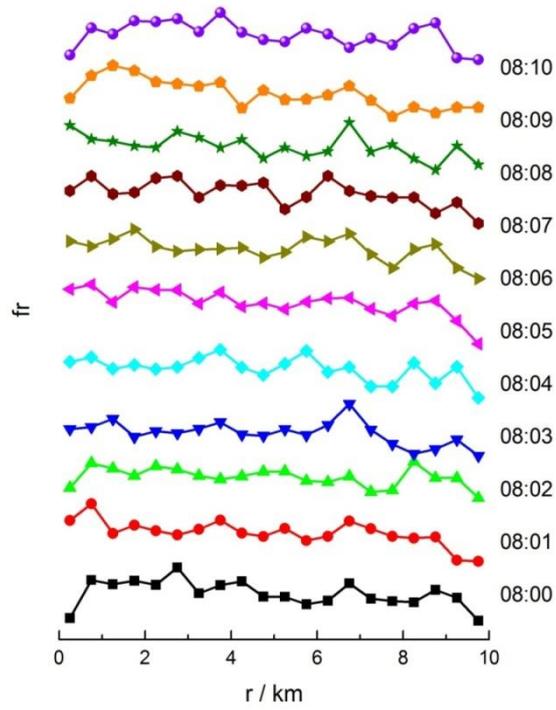

**Supplementary Figure 12:** $fr$ as a function of distance from centers in traffic network of Beijing.

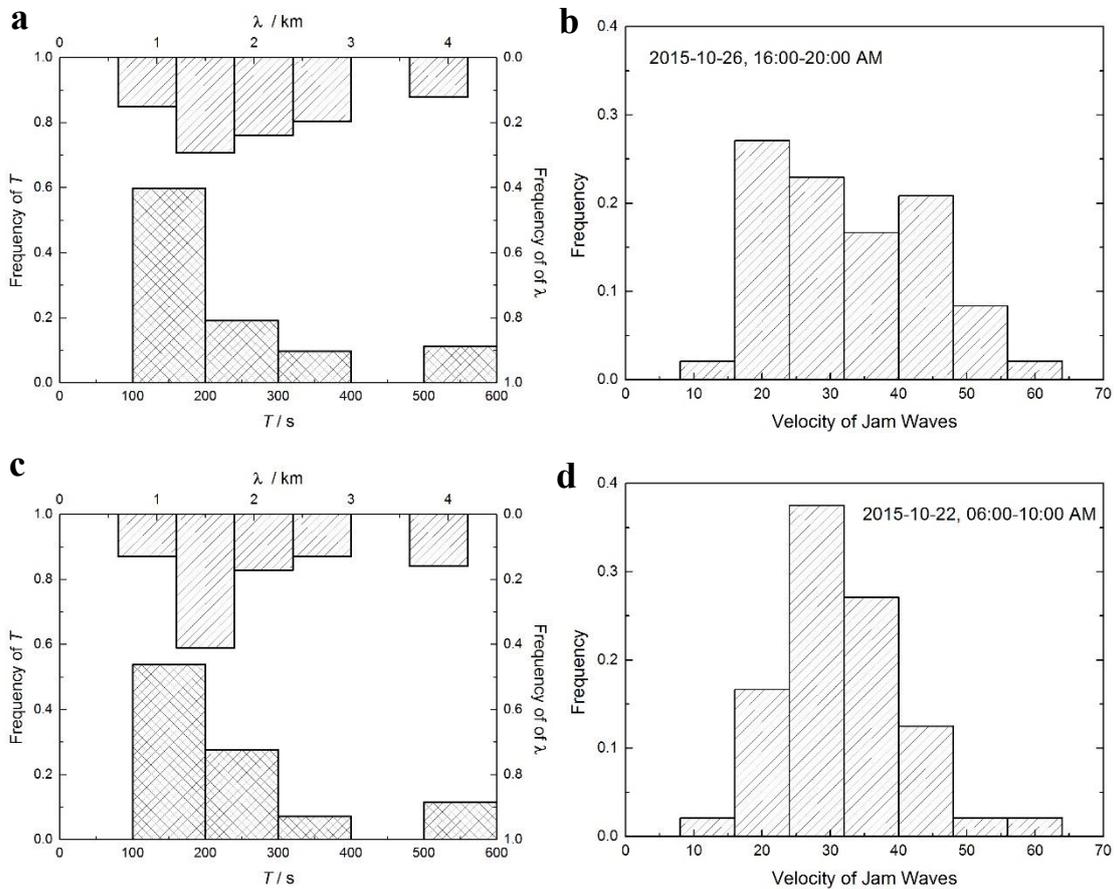

**Supplementary Figure 13: Velocities of jam waves in Beijing.** (a) and (c) Distributions of periods and wavelengths of jam waves during 16:00-20:00 on October 26[th], 2015 and 06:00-10:00 on October 22[th], 2015. (b) and (d) Distribution of velocities of jam waves in Beijing for the corresponding time periods in (a) and (b).

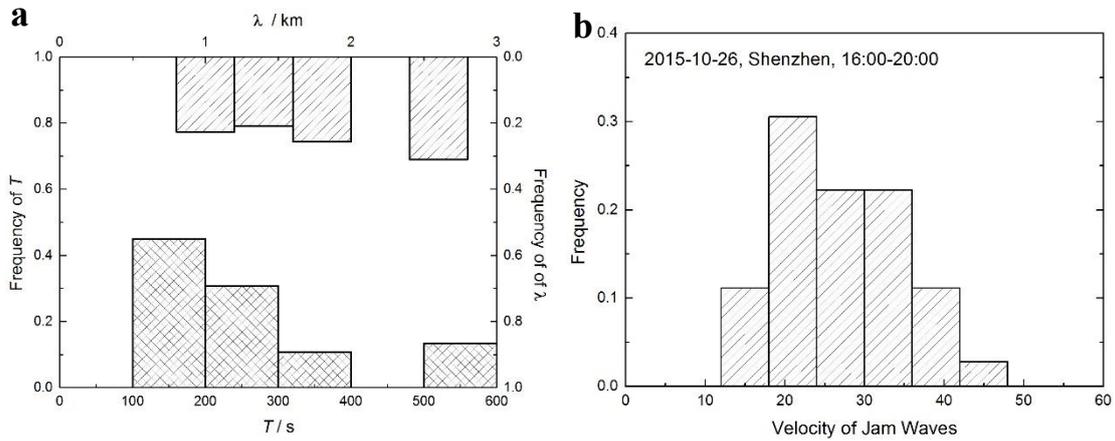

**Supplementary Figure 14: Velocities of jam waves in Shenzhen.** (a) Distributions of periods and wavelengths of jam waves during 16:00-20:00 on October 26[th], 2015. (b) Distribution of velocities of jam waves in Shenzhen for the corresponding time periods in (a).

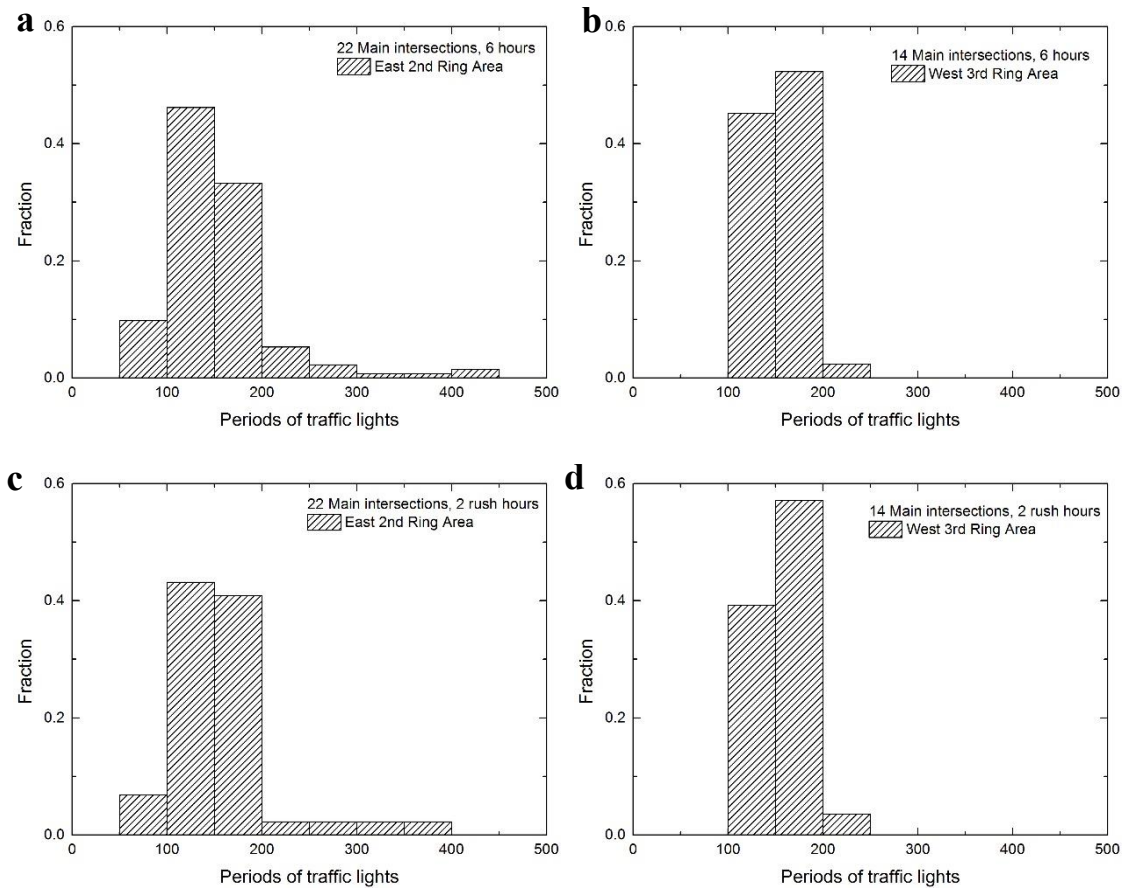

**Supplementary Figure 15: Distributions of periods of main traffic lights in two conventional congested area of Beijing.**

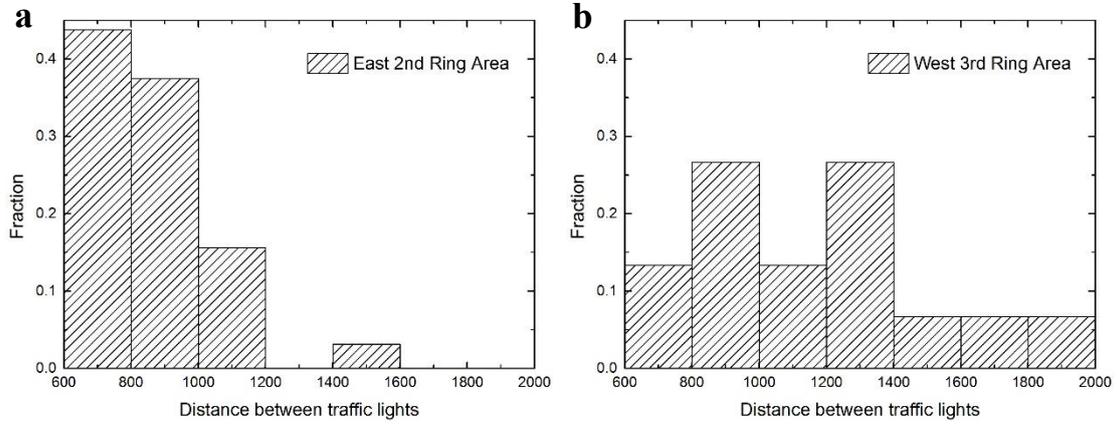

**Supplementary Figure 16: Distributions of distances between main traffic lights in two conventional congested area of Beijing.**